\documentclass[twocolumn,
superscriptaddress,
prl,
showpacs,
preprintnumbers,
amsmath,amssymb]{revtex4}

\usepackage{graphicx,epsf,epsfig,ulem}% Include figure files
\usepackage{epsfig}
\usepackage{bm}% bold math
\usepackage{subfigure}
\usepackage{color}

\begin{document}

\title{Coherent electron transport by adiabatic passage in an imperfect donor chain}% Force line breaks with \\

\author{Rajib Rahman}
\affiliation{Sandia National Laboratories, Albuquerque, NM 87185, USA}

\author{Richard P. Muller}
\affiliation{Sandia National Laboratories, Albuquerque, NM 87185, USA}

\author{James E. Levy}
\affiliation{Sandia National Laboratories, Albuquerque, NM 87185, USA}

\author{Malcolm S. Carroll}
\affiliation{Sandia National Laboratories, Albuquerque, NM 87185, USA}

\author{Gerhard Klimeck}
\affiliation{Network for Computational Nanotechnology, Purdue University, West Lafayette, IN 47907, USA}
%\affiliation{Jet Propulsion Laboratory, California Institute of Technology, Pasadena, CA 91109, USA}

\author{Andrew D. Greentree}
\affiliation{Center for Quantum Computer Technology, School of Physics, University of Melbourne, VIC 3010, Australia}

\author{Lloyd C. L. Hollenberg}
\affiliation{Center for Quantum Computer Technology, School of Physics, University of Melbourne, VIC 3010, Australia}

\date{\today}

\begin{abstract}
Coherent Tunneling Adiabatic Passage (CTAP) has been proposed as a long-range physical qubit transport mechanism in solid-state quantum computing architectures. Although the mechanism can be implemented in either a chain of quantum dots or donors, a 1D chain of donors in Si is of particular interest due to the natural confining potential of donors that can in principle help reduce the gate densities in solid-state quantum computing architectures. Using detailed atomistic modeling, we investigate CTAP in a more realistic triple donor system in the presence of inevitable fabrication imperfections. In particular, we investigate how an adiabatic pathway for CTAP is affected by donor misplacements, and propose schemes to correct for such errors. We also investigate the sensitivity of the adiabatic path to gate voltage fluctuations. %A large-scale atomistic tight-binding technique is used to capture donor states accurately taking into account the full bandstructure of the host, external gate potential, and device geometry. 
The tight-binding based atomistic treatment of straggle used here may benefit understanding of other donor nanostructures, such as donor-based charge and spin qubits. Finally, we derive an effective 3 $\times$ 3 model of CTAP that accurately resembles the voltage tuned lowest energy states of the multi-million atom tight-binding simulations, and provides a translation between intensive atomistic Hamiltonians and simplified effective Hamiltonians while retaining the relevant atomic-scale information. This method can help characterize multi-donor experimental structures quickly and accurately even in the presence of imperfections, overcoming some of the numeric intractabilities of finding optimal eigenstates for non-ideal donor placements.
\end{abstract}


\maketitle 

\section{I. Introduction}
Silicon quantum bits (qubits) are pursued due to the promise of long spin coherence times and the processing expertise of the semiconductor industry, that potentially could produce a quantum computer (QC). Over the past decade, many different Si qubits have been proposed. Some of these are based on nuclear \cite{Kane.nature.1998} or electronic spins \cite{Vrijen.pra.2000, Hill.prb.2005} of donors, while others are based on electronic spins in lithographically or electrostatically defined quantum dots \cite{Loss.pra.1998, Eriksson.NaturePhysics.2007}, and still others are based on the localized charge states \cite{Hollenberg.prb.2004.1} of confined electrons. 

\begin{figure}[htbp]
\center\epsfxsize=3in\epsfbox{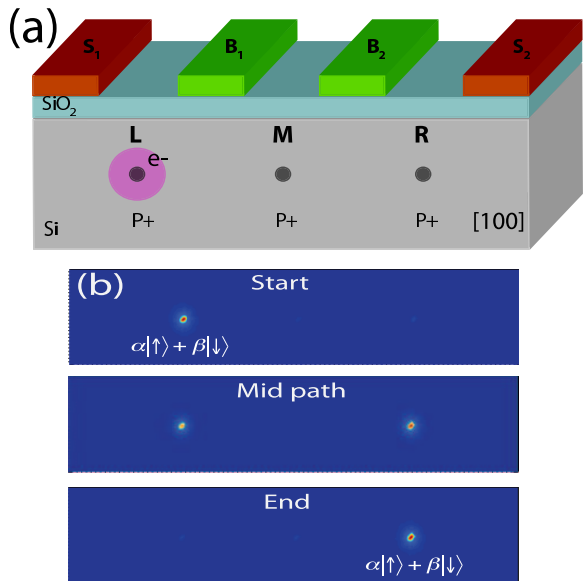}
\caption{(a) A three-donor CTAP device with surface (S) gates and barrier (B) gates. The system has 3 ionized donors (L = Left, M = Middle, R=Right) and 1 bound electron. %The gates can either detune the donor orbital states or can control the tunneling barriers between them. 
The device size is 60 nm $\times$ 30.4 nm $\times$ 30.4 nm. (b) The localization of the donor electron at 3 different voltage configurations in their stepped sequence. Although CTAP is a charge transport mechanism, in QC architectures it can be used to transport the spin superposition from the left to the right donor.} 
\vspace{-0.5cm}
\label{fig1}
\end{figure}

%The scalability of a QC architecture, however, does not merely depend on the ability to fabricate and link up many identical qubits, but on the ability to implement complex circuitry for qubit operation, measurement, and error correction such that individual qubits can be efficiently addressed by classical control mechanisms. Along these lines, 
QC architectures will benefit greatly if long-range coherent quantum transport schemes can be incorporated into the architecture for efficient transfer of the qubit state to different areas of the QC. In Ref \cite{Hollenberg.prb.2006}, a 2D bilinear array architecture for donor qubits was presented that utilized a novel non-local quantum transport mechanism called coherent tunneling adiabatic passage (CTAP) \cite{Greentree.prb.2004}. 

In essence, CTAP is a solid-state analogue of the well known STIRAP (stimulated raman adiabatic passage) protocol of quantum optics \cite{Vitanov.rpc.2001}. It is ideally suited to physically transporting quantum information across a chain of donors or quantum dots. STIRAP has been recently demonstrated in optical waveguides using photons \cite{Longhi.pre.2006, Longhi.jphys.2007, Longhi.prb.2007}. Apart from dot \cite{Greentree.prb.2004} and donor \cite{Hollenberg.prb.2006} based CTAP for quantum information transport, CTAP has also been proposed for transporting single atoms \cite{Eckert.pra.2004, Eckert.opt.2006} and Bose-Einstein condensates \cite{Graefe.pra.2006, Rab.pra.2008}. In previous works, comparisons have been made between the solid-state and quantum optics versions of CTAP \cite{Cole.prb.2008, Tom.pra.2009}. Recent works have explored alternate coupling schemes \cite{Jong.nanotech.2009} and CTAP based interferometry \cite{jong.prb.2010} in donor nanostructures. Since triple dot structures have already been fabricated \cite{Schroer.prb.2007, Grove.nanolett.2008, Gaudreau.prb.2009, Amaha.apl.2009, Pierre.apl.2009}, solid-state CTAP may not be far from reality.

In a system of tunnel-coupled donors or quantum dots, CTAP can be realized by modulating barrier gates (Fig. \ref{fig1}(a)) adiabatically in a counter intuitive pulsing sequence \cite{Greentree.prb.2004}. The protocol begins with an electron localized at one end of the chain in a superposition of its up and down spin states, and ends with the electron transported to the other end of the chain while retaining the quantum superposition in the spin basis (Fig. \ref{fig1}(b)). At all times during the transfer, there is negligible occupation probability of the electron at any point along the chain except for the ends. In effect, the protocol realizes certain pathways in the eigenspace through sensitive control of the molecular states of the donors or the dots. The method is expected to reduce gate densities relative to other quantum transport architectures in QCs, such as electron shuttling on the surface \cite{Skinner.prl.2003}. 

Bottom-up approaches have been successful in placing single donors in Si with a precision of 1 nm \cite{Schofield.prl.2003}. Recently, a 1D wire of donors has been fabricated by STM patterning single Phosphorus donors in Si \cite{Ruess.prb.2007.1}. Other recent STM patterned structures include a 2D delta doped layer of P donors for gating other nanostructures \cite{Ruess.prb.2007}, a single P donor surrounded by four leads \cite{Simmons.unpublished}, and a quantum dot formed by patterning a few donors \cite{Simmons.nnano.2010}. Top-down approaches by selective ion-implantation have also been successful in fabricating few donor devices \cite{Jamieson.apl.2005}. In Ref \cite{Donkelaar.njp.2010}, strategies to build a few donor CTAP device by ion-implantation was explored. Such experimental progress in donor placement motivates the examination of donor-CTAP's robustness to non-ideal placement of the donors and voltage fluctuations. %has brought CTAP in serious contention as an important qubit transport mechanism in solid state QCs.  

Since the inter-donor tunnel coupling is sensitive to the relative locations of the donors, even small placement errors of a lattice constant or less are likely to affect the adiabatic pathway in practical implementations of CTAP.  In fact, this donor straggle problem occurs in all donor qubits, whether spin or charge based. In the Kane qubit \cite{Kane.nature.1998}, the inter-donor exchange energy has been shown to oscillate with relative donor separations \cite{Koiller.prl.2001}. In a donor charge qubit, the symmetric-anti-symmetric gap ($\Delta_{SAS}$) also exhibits oscillations as a function of donor separation and orientation \cite{Hu.prb.2005}. The sensitivity of these parameters occur both due to the exponential fall-off of the envelope wavefunction and due to the rapidly oscillating Bloch functions resulting from the momentum states near the conduction band valleys of the host. A resolution to this problem is to experimentally characterize each donor qubit and their interactions \cite{Cole.pra.2005, Cole.pra.2006, Testolin.pra.2007}, and obtain voltage pulses that correct for these straggle effects. 

In a previous work \cite{Rahman.prb.2009}, the existence of an adiabatic pathway in an ideal triple donor chain was established using large scale atomistic tight-binding simulations. For the specific donor configuration in consideration, a set of voltages was found to implement a complete transfer sequence of the electron from one end of the chain to the other, thereby proving that a translation of this protocol from the quantum optics framework to a real solid-state nanostructure is valid and feasible. 

%CTAP is well-suited to materials like silicon, where the spin orbit interaction is small. Also, the transfer times have to be fast compared to orbital decoherence times, a condition easily met in the low temperature regime where phonon effects are strongly suppressed.

In this work, we investigate the robustness of the adiabatic path against imperfections such as donor straggle and gate voltage fluctuations. We also develop a general model that provides improved insight about the tunnel couplings and on-site energies of the multi-million atom system. The model can assist experiments to characterize donor nanostructures efficiently, and can help to design voltage pulses to correct for straggle. Although we restrict our attention to a triple donor chain, the methods and the results presented here can be easily translated to other 3D confined nanostructures in silicon and longer CTAP chains. 

%Although it may be easier to demonstrate CTAP in a triple dot structure for proof of concept studies, large scale architectures for quantum information transport may not benefit fully from a chain of dots due to the large number of gates needed to define each dot. The real advantage of CTAP will be realized in a chain of donors, where the naturally occurring Coulomb potential of the donors confine the electrons without the need for additional gates. In fact, 

We have employed atomistic tight-binding (TB) theory \cite{Slater.physrev.1954}, as this technique provides a highly accurate description of impurities in silicon \cite{Rahman.prl.2007, Park.prl.2009, Rahman.prb.2009.1, Rogge.NaturePhysics.2008} and allows the fast solution of million atom systems \cite{Klimeck.cmes.2002}. This method also treats the full band structure of the host, and provides a unified framework to treat realistic geometries, gate voltages, and disorder in an atomistic setting \cite{Klimeck.ted.2007}.   

This paper is organized as follows. In Section II, we describe the specifics of the nanostructure considered here, and introduce a solution for the adiabatic pathway with ideal donor placements \cite{Rahman.prb.2009}. Section III elaborates on the method of calculation. Section IV describes the results in detail. We show direct TB simulations of the effect of straggle and its corrections. We also show an effective $3 \times 3$ matrix model based on TB wavefunctions that can capture the effect of the gates and the details of the straggle.  

\section {II. Adiabatic path in an ideal triple donor CTAP device}

The CTAP device used in this work is shown in  Fig. \ref{fig1}(a). For consistency, we have used the same device as in Ref \cite{Rahman.prb.2009}, except displacing the donor positions to study effects of straggle. The structure consists of about 3.5 million Si atoms with one electron bound across three ionized P donors. Two symmetry gates and two barrier gates, each of 10 nm width, are placed above a 5 nm thick oxide layer on top of the Si lattice. Ideally, the barrier (B) gates modulate the tunnel barriers between the donors, whereas the symmetry (S) gates detune the energies of the end donors. The donors are buried 15 nm below the oxide, and are also placed 15 nm apart from each other in the [100] direction in the ideal case. The closeness of the gates in this particular device gives rise to significant crosstalk. %Control is likely to be easier in a more realistic device with 20-30 nm separation between the donors. 
Further details of the structure can be found in Ref \cite{Rahman.prb.2009}.

The three donor system can be conceptually understood with an effective 3 $\times$ 3 Hamiltonian,

\begin{equation} \label{eq:H0} 
\left( \begin{array}{ccc}
0 & t_{LM} & 0 \\
t_{LM} & 0 & t_{MR} \\
0 & t_{MR} & 0
\end{array} \right)
\end{equation}

\noindent
where $t_{LM}$ and $t_{MR}$ are the gate voltage tuned tunnel couplings between the donors $L$, $M$, and $R$. We assume a simplified case for this discussion such that the donor on-site energies are aligned, which can be practically achieved via tuning by the symmetry gates $S_1$ and $S_2$. The lowest states of this Hamiltonian are,

\begin{align}								
|\Psi_j \rangle =  \alpha_j |L \rangle + \beta_j |M \rangle + \gamma_j |R \rangle 
%&|\Psi_2 \rangle= \alpha_2 |L \rangle + \beta_2 |M \rangle + \gamma_2 |R \rangle \\  \nonumber \\
%&|\Psi_3 \rangle= \alpha_3 |L \rangle + \beta_3 |M \rangle + \gamma_3 |R \rangle
\end{align}

\noindent
where $\alpha_j$, $\beta_j$, and $\gamma_j$ are the coefficients of the donor states $|L\rangle$, $|M\rangle$, and $|R\rangle$, respectively for the eigenstate $j$. Here, $j$ runs from 1 to 3, with $j=1$ representing the ground state. $\alpha$, $\beta$, and $\gamma$ are each functions of $t_{LM}$ and $t_{MR}$. For ideal CTAP $\beta_2=0$ at all times.  
Notionally CTAP operates through maintaining the charge degree of freedom in the first excited state, $|\Psi_2 \rangle$, and adiabatically evolving $|\Psi_2 \rangle$, such that the wavefunction is transferred from $| L \rangle$ at the start to $ | R \rangle$ at the end through voltage tuning such that $\gamma_2=0$ at the start and $\alpha_2=0$ at the end. The first excited state $|\Psi_2 \rangle$ is protected from $\Psi_1 \rangle$ and $\Psi_3 \rangle$ by energy gaps $\Delta_{12}$ and $\Delta_{23}$, respectively. We will discuss later in this paper a method to project the multi-million atom Hamiltonian on to a 3 $\times$ 3 subspace, which can provide an accurate description of the full Hamiltonian for the relevant states. This model is one of the important contributions of this paper.

\begin{figure}[htbp]
\center\epsfxsize=3.4in\epsfbox{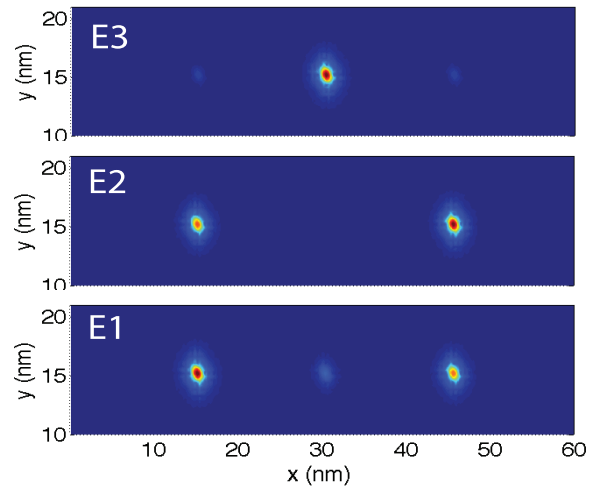}
\caption{(a) The lowest three eigenstates of the system near the mid-point of an adiabatic path for CTAP. These were obtained from tight-binding simulations of the device containing about 3.5 million atoms. The transfer takes effect through the %initially degenerate 
first excited state of the system (E2), in which the electron density diminishes at the center donor. This point in the adiabatic path was realized for the gate configuration $(V_{S1}, V_{B1}, V_{B2}, V_{S2})=(-0.01325, -0.11, -0.11, 0)$ V. %(b) Variation of the tunnel gaps with donor separations if the applied voltage is fixed at the configuration in (a).} %The wavefunction symmetries shown here are a characteristic of the adiabatic pathway for CTAP. By tuning gate biases, these wavefunctions can be modulated to transfer an electron from the left (L) to the right (R) donor with negligible occupation at the middle (M) donor.
} 
\vspace{-0.5cm}
\label{fig2}
\end{figure}

In Fig. \ref{fig2}, we show the molecular states of the triple donor system at the midpoint of the adiabatic transfer path for the CTAP protocol. %These states form at a gate configuration of $(V_{S1}, V_{B1}, V_{B2}, V_{S2})=(-0.01325, -0.11, -0.11, 0)$ V \cite{Rahman.prb.2009}. 
E1 represents the ground state, while E2 and E3 are the first and second excited states, respectively. CTAP transport takes effect through the state E2. %At the start of the protocol, the gates can be configured such that E2 has electron density only at the left donor. At the end, another gate configuration can localize the electron density only at the right donor. Between these two points, the gates can be pulsed such that the electron density gradually diminishes from the left donor and reappears on the right donor, without any significant population forming at the middle donor at any point in time. 
In an ideal scenario, the middle donor between the barrier gates can be replaced by a chain containing an odd number of donors, and the same transfer protocol can still hold, without the need for any additional gates \cite{Greentree.spie.2005}. 

In this work we will use the relative electron density in the center donor in the state E2 ($| \Psi_2 \rangle$) as a metric of the efficacy of the CTAP transfer. The justification for this metric directly comes from earlier works \cite{Greentree.prb.2004, Cole.prb.2008, Rab.pra.2008, Tom.pra.2009}. Adiabatic transfer is realized through the wavefunction symmetries described earlier in this section and also shown in Fig. \ref{fig2} for the midpoint of the adiabatic path. The vanishing center donor density in the state E2 ($| \Psi_2 \rangle$) is a defining feature of this adiabatic transfer. Any marked build up of center donor density in E2 indicates a deviation of the wavefunction symmetries required for CTAP, and is an indication that the transfer is becoming more non-adiabatic.

\begin{figure}[htbp]
\center\epsfxsize=3.2in\epsfbox{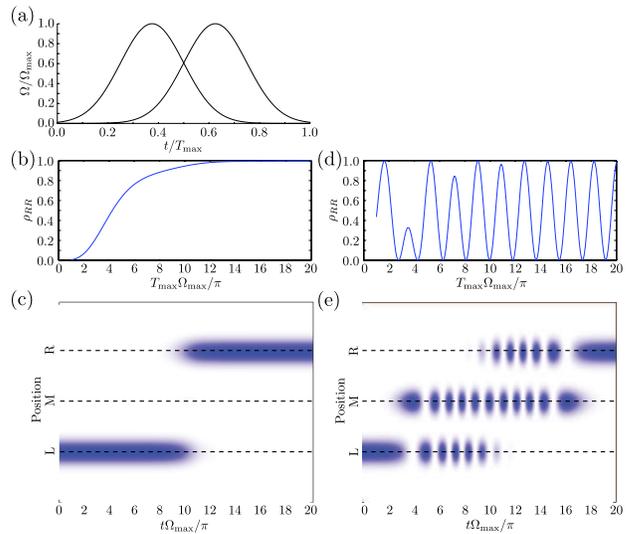}
\caption{Population evolution through CTAP sequence. (a) For these calculations we use a Gaussian pulse sequence.  For the intuitive direction, $t_{LM}$ is the leading pulse, followed by $t_{MR}$.  In the counter-intuitive direction, this order is reversed. (b) Final state population at donor R ($\rho_{RR}$) for increasing \emph{total} time of the pulse sequence in the counter-intuitive direction. Note the smooth, monotonic improvement in transfer fidelity with increasing time. (c) Population as a function of time through the CTAP protocol for total time $T_{\max} = 20 \pi/\Omega_{\max}$. Observe the smooth population transfer with negligable population at the central donor. (d) As (b) but for the intuitive pulse sequence.  Note the oscillating population, with the final state population depending sensitively on the total time. (e) As (c) for the intuitive direction, again showing that the population exhibits nonadiabatic oscillations through the transfer protocol.} 
\vspace{-0.5cm}
\label{fig3}
\end{figure}

In Fig. \ref{fig3}, we performed a time dependent analysis based on the effective 3 $\times$ 3 CTAP density matrix to show the relation between the donor densities in the adiabatic and non-adiabatic regimes. The time evolution of the densities was obtained from $d{\rho}/dt=-(i/{\hbar})[H, \rho]$. We ignored effects of decoherence for this analysis as it is beyond the scope of this work, and was considered in Ref \cite{Kamleitner.pra.2008} in the more general context of adiabatic information transport. To effect an electron transfer from donor $L$ to donor $R$, CTAP relies on a counter-intuitive pulsing (CIP) of the barrier gates, in which $B_2$ is pulsed before $B_1$. On the other hand, the more conventional sequential transport from donor R to M and then to L is non-adiabatic in nature, and can be realized with an intuitive pulsing (IP) scheme, in which $B_1$ is pulsed before $B_2$. Fig. \ref{fig3} shows the Gaussian pulses (in (a)) and the evolution of the donor densities in time (in (b), (c), (d), and (e)). The IP sequence produces an oscillating donor density in the end donor $R$ (Fig. \ref{fig3}(d)) and significant population in $M$ (Fig. \ref{fig3}(e)) which oscillates in time. This pulse does not guarantee a robust final transfer to the end donor. The CIP scheme evolves the system adiabatically as the density in $R$ (Fig. \ref{fig3}(b)) smoothly increases to 1 with vanishing density at $M$ (Fig. \ref{fig3}(c)) at all times, making CTAP a robust transport method.

%are shown in Fig. Calculations of the electron density at right donor $R$ as as a function of time (Fig. \ref{fig4} of Ref \cite{Greentree.prb.2004}) show that the counter-intuitive pulsing scheme produces a negligible center density at any time during the transfer, and offers a fast and smooth transfer to the end donor $R$. In the intuitive pulsing scheme, significant center density builds up in the midst of the transfer, and this relates to an oscillatory density at the end donor over a significantly longer time step. The intuitive pulsing scheme effects the electron transport through the ground state, whereas the counter-intutive pulsing scheme effects transport through the 1st excited state. CTAP transfer is therefore a fast and robust method of coherent transport, and is directly related to a vanishing electron density at the center donor.

The build up of center donor density in E2 represents a deviation from ideal CTAP. Imperfections such as decoherence, straggle, and voltage noise can enhance this center density, taking the system into a more non-adiabatic transfer regime. The center donor density is thus a degree of measure of the non-ideality of a CTAP transfer, as suggested also in Refs \cite{Cole.prb.2008, Rab.pra.2008, Tom.pra.2009}.

\section {III. Method} 

Since in principle CTAP is an adiabatic problem, we analyze it by solving the time independent Schr\"odinger equation at different bias points. We have employed the 10 band $sp^{3}d^{5}s^{*}$ tight-binding model with nearest-neighbour interactions. The model parameters were optimized by a genetic algorithm with appropriate constraints to reproduce the important features of the bulk bandstructure of the host \cite{Slater.physrev.1954}. The model parameters have been well-established in the literature \cite{ Boykin.prb.2004, Klimeck.cmes.2002}, and calculations performed with these parameters have been verified against experimental measurements in a number of works \cite{Klimeck.ted.2007, Rahman.prl.2007, Rogge.NaturePhysics.2008, Park.prl.2009, Neerav.apl.2007}. 
 
Each P donor was modeled by a Coulomb potential screened by the dielectric constant of Si. At the donor site, a cut-off potential $\textrm{U}_0$ was used, and its value optimized so that the ground state binding energy of -45.6 meV was obtained for a donor in bulk Si. In this TB model, the valley-orbit interaction that lifts the six-fold degeneracy of the donor ground state is inherently included \cite{Shaikh.encyclopedia.2008}. 
 
The electrostatic gate potential was obtained from a commercial Poisson solver \cite{ISE.TCAD} for a single gate, and the potential for the three gates was treated as a superposition of the single gate solution. Although the voltages presented here will have some offsets from the realistic case due to this, the basic trends and analysis presented here are general, and this approach reduced the computational resources necessary to complete the calculations in a tractable time. 
 
The net potential was then interpolated onto the atomistic grid for the tight-binding simulations. Closed boundary conditions with a model of dangling bond passivation was used to model the interfaces \cite{Lee.prb.2004}. The full Hamiltonian of about 3.5 million atoms including the four gate potentials was solved by parallel Lanczos and block Lanczos algorithms to capture the relevant eigenvalues and wave functions. Typical computation time for a single time-step of the protocol over 6 states was 7 hours on 40 processors {\cite{nanohub.note}}. 

\section{IV. Results and Discussions}

\subsection{A. Binding energy of $P_{3}^{2+}$ with straggle under zero gate bias}

\begin{figure}[htbp]
\center\epsfxsize=3.4in\epsfbox{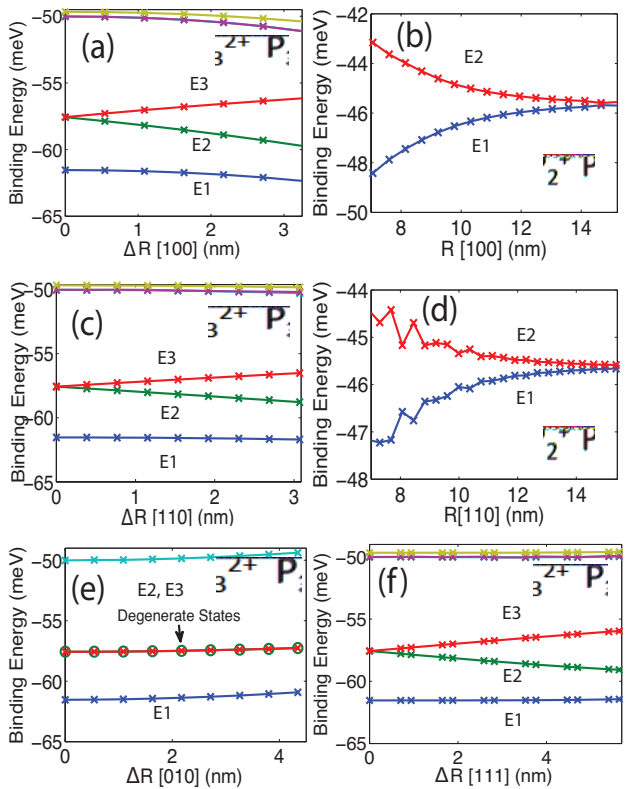}
\caption{The lowest few states of the triple donor system at zero gate bias. The binding energy is expressed relative to the conduction band minima of Si. $\Delta{R}$ is the distance the middle donor is moved towards the right donor from the ideal (equidistant) case of 15.2 nm separation. The states for $\Delta R$ along (a) [100] (x), (c) [110] (x=y), (e) [010] (y), and (f) [111] (x=y=z).%The lowest donor states for a displacement of the middle donor in the [110] plane. %The sensitivity of the first two excited states to small displacements of the middle donor indicates that the precisely tuned adiabatic pathway is likely to be affected by donor straggle. 
%Plots (c) and (d) are for donor straggle perpendicular to the chain (along [010]) and along [111] respectively.
Plots (b) and (d) compare the lowest two (bonding and anti-bonding) states of a donor charge qubit ($P_{2}^{+}$) for donor separations along [100] (x) and [110] (x=y) respectively.} 
\vspace{-0.5cm}
\label{fig4}
\end{figure}

In a single P donor, the lowest manifold consists of 6 1s type orbital states, arising from the six-fold degenerate conduction band valleys of Si. Due to the valley-orbit interaction caused by central cell effects \cite{Ramdas.progphys.1981}, the six 1s states are split into an orbital singlet $A_1$, an orbital triplet $T_2$, and an orbital doublet $E$ \cite{Kohn.physrev.1955}. The $A_1$ state for a P donor is separated by 11.7 meV from the $T_2$ states. In a triple donor chain with donor separations much larger than the donor Bohr radii, the E1, E2, and E3 states arise from linear combinations of the $A_1$ states of each impurity. The 12 meV gap between the $A_1$ and the $T_2$ states of a single donor also gives rise to a significant energy gap between the lowest 3 states and the higher manifold of the $P_{3}^{2+}$ molecule, which is a desirable condition for efficient CTAP transfer. %The calculated gaps with donor separations are shown in Fig. \ref{fig2}(b). 
Translation of this concept to quantum dots will be straightforward only if the valley splitting is large such that the lowest three states involved in CTAP are sufficiently isolated in energy from the valley split manifold.  

In Fig. \ref{fig4}, we show the lowest energy states of the $P_{3}^{2+}$ molecule at zero gate bias as a function of non-ideal placement of the middle donor. $\Delta{R}$ represents the distance the middle donor is displaced towards the right donor relative to its ideal (equidistant) location. For example, in Fig. \ref{fig4}(a), $\Delta{R}=0.543$ nm means that the middle donor is displaced by one lattice constant $a_0$ ($0.543$ nm) along the donor chain. 

The states E2 and E3, which are degenerate at $\Delta{R}=0$, split due to asymmetric tunnel coupling to the left and right donors, as shown in Fig. \ref{fig4}(a), 4(b) and 4(c). The order of meV splitting for the range of donor separations considered here is comparable to the order of gate voltage modulation for electron transfer. Hence, donor straggle is likely to affect the adiabatic transfer. For large donor misplacements of $10a_0$ or more, there is the added problem that the higher manifold of states move toward the 2nd excited state E3, an undesirable condition for efficient transfer. This represents a potentially significant challenge to implement buried CTAP. 

Fig. \ref{fig4}(c) and 4(f) are for donor straggle along [110] and [111] directions respectively relative to the ideal chain in [100]. The striking feature is that the variations are almost smooth, compared to the oscillations in symmetric-antisymmetric energy states ($\Delta_{SAS}$) for a charge qubit of $P_{2}^{+}$ molecule shown in Fig. \ref{fig4}(d) and also in Ref \cite{Hu.prb.2005}. In the triple donor system, the oscillations in energy states with donor separations seem to be somewhat mitigated. This indicates that relatively smooth voltage corrections are possible. 

Fig. \ref{fig4}(e) shows the effect of straggle on the donor spectrum for a displacement perpendicular to the chain along [010] direction. In this case, the spectrum is less affected as the middle donor remains equidistant from the left and right donors. The states E2 and E3 remain degenerate throughout as expected.  

\begin{figure}[htbp]
\center\epsfxsize=3.4in\epsfbox{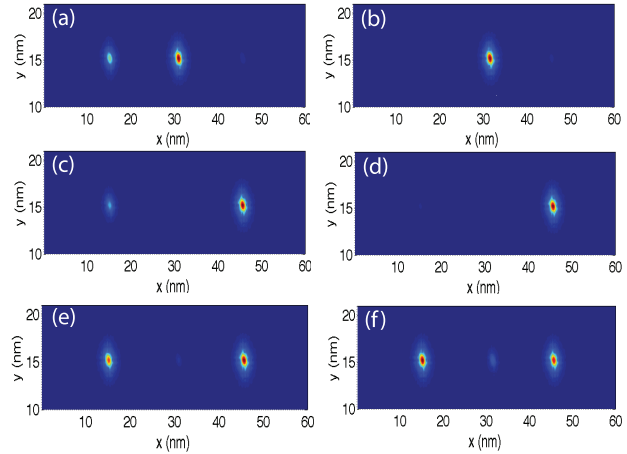}
\caption{Plots of the 1st excited state (E2) electron density for donor straggle of (a) $\Delta R=1a_0 [100] (x)$ ($a_0=0.543$ nm being the lattice constant of Si), (b) $\Delta R=2a_0 [100] (x)$, (c) $\Delta R=1a_0 [010] (y)$, (d) $\Delta R=2a_0 [010] (y)$, (e) $\Delta R=1a_0 [100] (x)$ with a corrective bias of $(-0.01325, -0.11, -0.11, -0.0084)$ V, and (f) $\Delta R=2a_0 [100] (x)$ with a corrective bias of $(-0.01325, -0.11, -0.11, -0.0170)$ V. %For the ideal case of equal donor separation, the electron density was negligible at the center donor. The central donor density is used as a metric of the effectiveness of the CTAP transfer. 
Cases (e) and (f) are the same as cases (a) and (b) respectively, except a corrective voltage is applied to the $S_2$ gate. %show the 1st excited state for cases (a) and (b) respectively, if corrective voltages are applied to the $S_2$ gate. Case (e) is under a bias $(-0.01325, -0.11, -0.11, -0.0084)$ V, whereas case (f) is under a bias $(-0.01325, -0.11, -0.11, -0.0170)$ V.
} 
\vspace{-0.5cm}
\label{fig5}
\end{figure}

\subsection{B. Effect of straggle in donor position on the adiabatic path}

To investigate the effect of straggle more exactly, we have chosen the midpoint of the adiabatic path shown in Fig. \ref{fig2} for the ideal case. Under the same gate bias, we have displaced the middle donor to neighboring lattice sites, and investigated the effect on the CTAP states. In particular, we compared the 1st excited state E2 obtained in this manner with the ideal E2 shown in Fig. \ref{fig2}.

In Fig. \ref{fig5}(a) and 5(b), the middle donor is displaced by $1a_0$ and $2a_0$ respectively towards the right donor (in [100] (x) direction). It is observed that the population becomes dominant at the center donor. If the donor is displaced perpendicular to the chain ([010] (y) direction) as in 5(c) and 5(d), the center donor density is still negligible. This suggests an intuition that straggle effects are maximum when donors are misplaced along the chain, since the tunnel coupling is asymmetrically affected the most in such cases. 

In Fig. \ref{fig5}(e) and 5(d), we have found voltages that can be applied to the $S_2$ gate to correct for the $1a_0$ and $2a_0$ straggle effects of 5(a) and 5(b), respectively, even in the presence of considerable perturbation from other neighboring gates. This means that external macroscopic voltages can be used to compensate for straggle effects due to single atom placement errors, which is a central point of this paper.

\begin{figure}[htbp]
\center\epsfxsize=3.4in\epsfbox{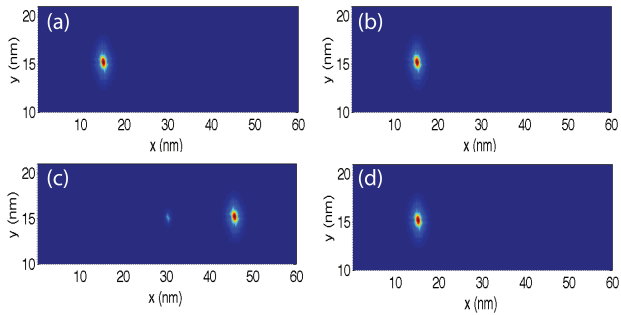}
\caption{Plots of the 1st excited state (E2) electron density for donor straggle of (a) $\Delta R=0.5a_0 [110]$, (b) $\Delta R=1a_0 [110]$, (c) $\Delta R=-0.25a_0 [111]$, and (d) $\Delta R=0.75a_0 [111]$.} 
\vspace{-0.5cm}
\label{fig6}
\end{figure}

Similarly, in Fig. \ref{fig6}, we investigated the effect of donor displacement along [110] in (a) and (b), and along [111] in (c) and (d). These directions are expected to be more susceptible to rapid spatial oscillations of the host Bloch functions, as seen in two-donor charge qubits \cite{Hu.prb.2005}. However, it is found that the center population is still somewhat negligible, which suggests that the effect of the crystal momentum states is not significant in this triple donor system. 

\subsection{C. Sensitivity of the protocol to gate voltage fluctuations}

\begin{figure}[htbp]
\center\epsfxsize=3.0in\epsfbox{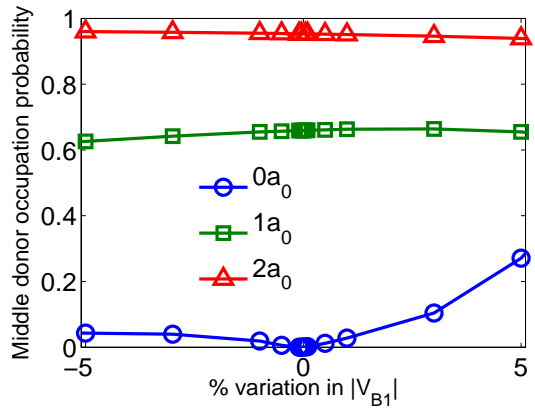}
\caption{Sensitivity of the adiabatic pathway to fluctuations in the barrier gate voltage $V_{B1}$. The plot is for the mid point of the path with $(-0.01325, -0.11, -0.11, 0)$ V for which the center donor population is $5.4e-4$ out of 1. The center donor density is plotted as a function of percentage variation of $|V_{B1}|$ for a range of -5  to +5 percent for the ideal case as well as straggle of 1 and 2 lattice constants along the chain.} 
\vspace{-0.5cm}
\label{fig7}
\end{figure}

The transfer efficiency of the CTAP protocol
in realistic systems may be susceptible to voltage
noise. In particular, voltage fluctuations in the
system can originate from charge traps or from
small fluctuations in the gate pulses. Such variations
are likely to affect both the tunnel barriers
and the detuned energies of the donors,
thereby introducing a finite population in the center
donor.

In Fig. \ref{fig7}, we have again used the center donor population
as a measure of the effectiveness of the transfer, and
investigated the robustness of the adiabatic path
to voltage fluctuations. For simplicity, we have
modeled the voltage noise by perturbing $V_{B1}$ by a percentage of its value at the mid point of the adiabatic path.  

Although Fig. \ref{fig7} shows that such
voltage offsets indeed introduce some population
at the center, it is to be noted that center donor density remains considerably smaller for the more negative bias range. At -5 percent change in $V_{B1}$, the population at the center is still about 0.04 out of 1. However, the center donor density is very sensitive to the more positive bias range. The transfer efficiency is seriously hampered after +2-3 percent fluctuation in $V_{B1}$.  

This asymmetry in sensitivity occurs due to the ordering of the CTAP states. As shown in Fig. \ref{fig2}, the state with the most central donor density is the 2nd excited state (E3) of the system. However, at zero gate bias, this state is the ground state of $P_{3}^{2+}$ due to strong tunnel coupling between the donors. Negative barrier gate biases were required to rearrange the states to realize the adiabatic path. Hence, more positive gate biases causes this state to be either the 1st excited state or ground state of the system, and thus breaks down the adiabatic path. Negative bias fluctuations, however, keep the ordering of the states intact, and have less effect on the path. 

\subsection{D. Effective 3 $\times$ 3 model developed from tight-binding}

Problems such as CTAP and STIRAP are well-described by small scale effective models, portrayed as a 3 $\times$ 3 matrix in the 3-donor CTAP, for example. However, a complete and realistic  description of the system has to include atomic scale effects from a method such as tight-binding. In this section, we describe a procedure to translate a complex atomistic system into a simpler 3 $\times$ 3 matrix model that retains the relevant atomic scale information. Such a model is general beyond CTAP, and helps to bridge the gap between intensive atomistic theories and reduced order envelope function methods.

Although atomistic TB simulations capture the realistic features of a CTAP device and are useful to guide CTAP experiments of the future, a typical experiment will involve scanning over a large bias range with any number of possibilities for straggled donor positions. Thousands of large scale TB simulations to understand the device will be impractical and time-consuming. For this purpose, we have developed an effective 3 $\times$ 3 model constructed from a few TB simulations of the system. In the perturbative regime, in which gate voltages cause small changes in the $A_1$ state of each impurity, this model can provide the same information as the actual TB solution of the full atomistic Hamiltonian. The advantage of this model is that a large number of voltages and straggled positions can be explored very rapidly, and correction procedures can be determined quickly. By imposing symmetry conditions and combining numerical parameterization and analytic solutions, we can even find sets of gate biases for various points of the adiabatic path, as we demonstrate below.    

%The concept of CTAP was originally described by a simple effective model. 
Assuming three different donor sites, and a wavefunction localized in each donor, we can use a 3 $\times$ 3 Hamiltonian describing the system in this 3-state basis. This Hamiltonian $H_{\rm eff}$ is of the form,
 %The diagonal terms of this matrix represent on-site energies of the donors, while off-diagonal terms represent coupling between two donors. 

%\begin{displaymath}
\begin{equation} \label{eq:H1} 
H_{\rm eff} =
\left( \begin{array}{ccc}
E_L & t_{LM} & t_{LR} \\
t_{LM}* & E_M & t_{MR} \\
t_{LR}* & t_{MR}* & E_R
\end{array} \right)
\end{equation}
%\end{displaymath}

\noindent
where $E_i$ is the on-site energy of the $i$-th impurity, and $t_{ij}$ is the tunneling matrix element from impurity $i$ to impurity $j$. %We can further simplify the system by assuming the ground state of the donors are aligned in energy so that $E_L = E_M = E_R$, and arbitrarily set the eigenvalues to 0. %$E_1 = 0$, and $E_2 - E_1 = \Delta$. 
%We can also assume only nearest donor coupling by setting $t_{LR} = 0$. 
If the on-site energies of the donors are aligned closely, the tunneling elements $t_{LM}$ and $t_{MR}$ can be switched alternately from on to off adiabatically to effect the electron transfer \cite{Rahman.prb.2009}. %A detailed description of this model can be found in Ref \cite{Rahman.prb.2009}.

The idea is to make this model more realistic by calculating the matrix elements from atomistic TB simulations. The TB Hamiltonian for a single donor under zero gate bias is given by, $H_i=H_0+V_i(R_i)$, where $H_0$ is the crystal Hamiltonian, and $V_i$ is the core corrected Coulomb potential of impurity $i={L,M,R}$ located at $R_i = {R_L, R_M, R_R}$. If we solve for $H_i$ for each of the three donors separately, we can obtain a set of donor states $\lbrace \psi_{ji} \rbrace$, where $j$ is the state index, and $i$ is the impurity index. In this case, we restrict our attention to the donor ground state only. So, $j=1$, and $i=L, M, R$, and the basis set $\phi=\lbrace \psi_{1L}, \psi_{1M}, \psi_{1R} \rbrace$. The full TB Hamiltonian of the CTAP device is,

\begin{equation} \label{eq:H2} 
\begin{array}{c}
H_T=H_0+V_L(R_L)+V_M(R_M)+V_R(R_R)\\
+ V_G(V_{S1}, V_{B1}, V_{B2}, V_{S2})
\end{array}
\end{equation}
 
The matrix elements of $H_{\rm eff}$ is then given by, $[H_{\rm eff}]_{lk}=\langle \psi_{1l} | H_T | \psi_{1k} \rangle$. Whereas the full solution of $H_T$ is time intensive, requiring many matrix vector multiplications in Lanczos, it is fairly fast to evaluate each matrix element of the 3 $\times$ 3 model as it involves one matrix vector multiplication between the TB Hamiltonian and a vector, and one inner product between two vectors. Since the basis set $\Phi$ is non-orthogonal, the 3 $\times$ 3 problem to be solved is of the form, $H_{\rm eff} \phi=ES\phi$, where $S$ is the overlap matrix with elements $S_{lk}=\langle \psi_{1l} | \psi_{1k} \rangle$. The eigen solution of the $H_{\rm eff}$ gives the molecular states of the CTAP device in the basis of the $A_1$ states of the three donors. It is to be noted that by simply including the donor excited states in the basis (for $j>1$), the model can be made more comprehensive and applicable to other donor related problems as well. It is also fairly easy to extend this method to any number of donors.

\begin{figure}[htbp]
\center\epsfxsize=3.4in\epsfbox{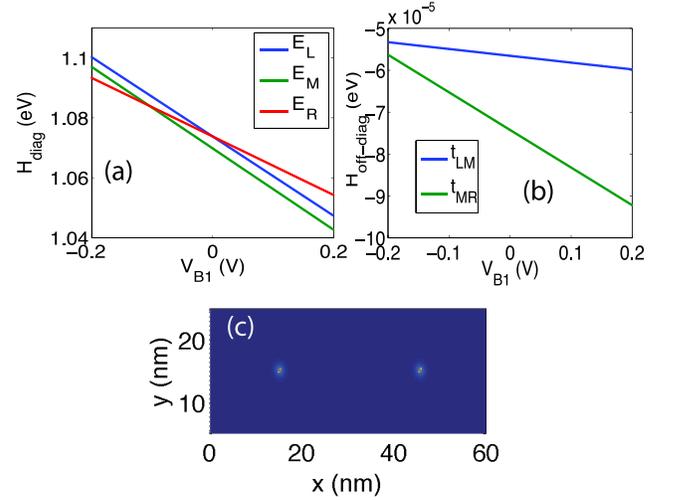}
\caption{ (a) Diagonal elements ($E_L$, $E_M$, $E_R$) of the 3 $\times$ 3 matrix as a function of $V_{B1}$. The other gates are set to zero. The donor M is equidistant from L and R. (b) The off-diagonal elements ($t_{LM}$ and $t_{MR}$) with $V_{B1}$. (c) The state E2 obtained in TB from the bias set $(0, -0.1128, -0.0903, 0)$ V. %The voltages were obtained from the parametrized 3 $\times$ 3 model.
} 
\label{fig8}
\vspace{-0.3cm}\end{figure}

Our approach at this point is to find the functional dependencies of the matrix elements on the four gate biases $V_G$, and atomic straggle as captured in the donor potential $V_i(R_i)$. In our case, we were able to perform linear fits of all the elements to the gate biases, with the function,

\begin{equation} \label{eq:H3} 
[H_{\rm eff}]_{lk} = a_1V_{S1}+a_2V_{B1}+a_3V_{B2}+a_4V_{S2}+a_5
\end{equation}
 
\noindent
where the coefficients $a_i$ are constants determined from the fit. Fig. \ref{fig8}(a) shows some diagonal elements as a function of $V_{B1}$, while 8(b) shows the same for the off-diagonal elements. 
 
Once this parameterized 3 $\times$ 3 is obtained, we verify that the model is capable of deducing the voltages for the adiabatic path. To obtain the midpoint of the path, we impose the conditions $t_{LM}=t_{MR}$, $E_L=E_M=E_R$, and $V_{S1}=V_{S2}=0$, and obtain the voltages, $V_{B1}=-0.1128$ V and $V_{B2}=-0.0903$ V from the matrix elements of $H_{\rm eff}$. These voltages are offset minutely from our brute force simulations by $0.01325$ V, $0.0028$ V, and $0.0197$ V for $V_{S1}$, $V_{B1}$, and $V_{B2}$, respectively. Performing the actual TB simulations with these voltages, we are able to obtain the midpoint of the adiabatic path, as shown by the wavefunction symmetry in Fig. \ref{fig8}(c).  In fact, we obtained an improvement over our previous voltages, of using only two gates instead of three. Therefore, the model is helpful for rapid evaluation of straggle effects. %provides a very quick way to obtain voltage sets for the adiabatic path, and can be used to guide experiments.

\section{V. Conclusion}

In this work, we investigated the effect of imperfections on coherent electron transport by adiabatic passage in a test triple donor system. Donor misplacements can severely hamper the adiabatic transfer path if the misplacements are along the chain, but can be retuned in theory, which is a central conclusion of this paper. We also performed a study of the sensitivity of the path to gate voltage fluctuations showing non-negligible sensitivity. We developed a quick and easy way to bridge between large scale tight-binding simulations and possible CTAP experiments. The effective 3 $\times$ 3 model that we developed from TB simulations for this purpose also unites the atomic scale precision of TB with intuitive simplicity of a toy model. The model in conjunction with tight-binding can help guide experiments even in the presence of imperfections such as donor straggle and gate voltage fluctuations. Furthermore, the model provides a general translation of intensive atomic scale calculations to simplified models used to describe problems in many solid-state and quantum-optics applications.  

\begin{acknowledgments}
Acknowledgments: This work was supported by the Laboratory Directed Research and Development program at Sandia National Labortories and by the National Security Agency Laboratory for Physical Sciences under contract number EAO-09-0000049393. Sandia is a multiprogram laboratory operated by Sandia Corporation, a Lockheed Martin Company, for the United States Department of Energy under Contract No. DE-AC04-94AL85000. Part of the development of NEMO-3D was initially performed at JPL,
Caltech under a contract with NASA. NCN/nanohub.org
computational resources were used. This work was supported by the Australian Research Council, NSA and ARO
(contract number W911NF-04-1-0290). A.D.G. and L.C.L.H. acknowldege the Australian Research Council for financial support (Projects No.
DP0880466 and No. DP0770715, respectively). R.R. also acknowledges discussions with T. Gurrieri and R. Young of SNL. 
\end{acknowledgments}

Electronic address: rrahman@sandia.gov, rmuller@sandia.gov


\vspace{-0.5cm}

\begin{thebibliography}{100}  %\setlength{\itemsep}{-0.5mm}

%\bibitem{Nielsen.book.2000} M. A. Nielsen and I. L. Chuang, {\it{Quantum Computation and Quantum Information}} (Cambridge Univ. Press, Cambridge, U. K., 2000).

\bibitem{Kane.nature.1998} B. E. Kane, Nature, {\bf{393}}, 133 (1998).

\bibitem{Vrijen.pra.2000} R. Vrijen, Eli Yablonovitch, K. Wang, H. W. Jiang, A. Balandin, V. Roychowdhury, T. Mor, and D. Divincenzo, Phys. Rev. A {\bf{62}}, 012306 (2000).

\bibitem{Hill.prb.2005} C. D. Hill, L. C. L. Hollenberg, A. G. Fowler, C. J. Wellard, A. D. Greentree, and H. -S. Goan, Phys. Rev. B {\bf{72}}, 045350 (2005). 


 % 2DEG QC
 \bibitem{Loss.pra.1998} D. Loss and D. P. Vincenzo, Phys. Rev. A {\bf{57}}, 120 (1998).

% Quantm wells
 \bibitem{Eriksson.NaturePhysics.2007} S. Goswami, K. A. Slinker, M. Friesen, L. M. McGuire, J. L. Truitt, C. Tahan, L. J. Klein, J. O. Chu, P. M. Mooney, D. W. Van Der Weide, R. Joynt, S. N. Coppersmith, and M. A. Eriksson, Nature Physics {\bf{3}}, 41 (2007).
 
 \bibitem{Hollenberg.prb.2004.1} L. C. L. Hollenberg, A. S. Dzurak, C. Wellard, A. R. Hamilton, D. J. Reilly, G. J. Milburn, and R. G. Clark, Phys. Rev. B {\bf{69}}, 113301 (2004).
 

\bibitem{Hollenberg.prb.2006} L. C. L. Hollenberg, A. D. Greentree, A. G. Fowler, and C. J. Wellard, Phys. Rev. B {\bf{74}}, 045311 (2006).

\bibitem{Greentree.prb.2004} A. D. Greentree, Jared H. Cole, A. R. Hamilton, and L. C. L. Hollenberg, Phys. Rev. B {\bf{70}}, 235317 (2004).

\bibitem{Vitanov.rpc.2001} V. Vitanov, T. Halfmann, B. Shore, and K. Bergmann, Annu. Rev. Phys. Chem. {\bf{52}}, 763 (2001).

% CTAP in optical fiber

\bibitem{Longhi.pre.2006} S. Longhi, Phys. Rev. E {\bf{73}}, 026607 (2006).

\bibitem{Longhi.jphys.2007} S. Longhi, J. Phys. B {\bf{40}}, F189 (2007).

\bibitem{Longhi.prb.2007} S. Longhi, G. Della Valle, M. Ornigotti, and P. Laporta, Phys. Rev. B {\bf{76}}, 201101(R) (2007).

%Transport of atoms

\bibitem{Eckert.pra.2004} K. Eckert, M. Lewenstein, R. Corbalan, G. Birkl, W. Ertmer, and J. Mompart, Phys. Rev. A {\bf{70}}, 023606 (2004).

\bibitem{Eckert.opt.2006} K. Eckert, J. Mompart, R. Corbalan, M. Lewenstein, and G. Birkl, Opt. Commun. B {\bf{264}}, 264 (2006).

% CTAP in Bose-Einstein Condensates

\bibitem{Graefe.pra.2006} E. M. Graefe, H. J. Korsch, and D. Whitthaut, Phys. Rev. A {\bf{73}}, 013617 (2006).

\bibitem{Rab.pra.2008} M. Rab, J. H. Cole, N. G. Parker, A. D. Greentree, L. C. L. Hollenberg, and A. M. Martin, Phys. Rev. A {\bf{77}}, 061602 (2008).

%CTAP papers

\bibitem{Cole.prb.2008} J. H. Cole, A. D. Greentree, L. C. L. Hollenberg, and S. Das Sarma, Phys. Rev. B {\bf{77}}, 235418 (2008).

\bibitem{Tom.pra.2009} Tom\'a\^s Opatrn\'y and Kunal K. Das, Phys. Rev. A {\bf{79}}, 012113 (2009).

\bibitem{Jong.nanotech.2009}L. M. Jong, A. D. Greentree, V. I. Conrad, L. C. L. Hollenberg, and D. Jamieson, Nanotechnology {\bf{20}}, 405402 (2009).

\bibitem{jong.prb.2010} L. M. Jong, and A. D. Greentree, Phys. Rev. B {\bf{81}}, 035311 (2010).

\bibitem{Schroer.prb.2007} D. Schršer, A. D. Greentree, L. Gadreau, K. Eberl, L. C. L. Hollenberg, J. P. Kotthaus, and S. Ludwig, Phys. Rev. B {\bf{76}}, 075306 (2007).

\bibitem{Grove.nanolett.2008} K. Grove-Rasmussen, H. I. J¿rgensen, T. Hayashi, P. E. Lindelof, and T. Fujisawa, Nanoletters {\bf{8}}, 1055-60 (2008).

\bibitem{Gaudreau.prb.2009} L. Gaudreau, A. S. Sachrajda, S. Studenikin, A. Kam, F. Delgado, Y. P. Shim, M. Korkusinski, and P. Hawrylak, Phys. Rev. B {\bf{80}}, 075415 (2009).

\bibitem{Amaha.apl.2009} S. Amaha, T. Hatano, T. Kubo, S. Teraoka, Y. Tokura, S. Tarucha, and D. G. Austing, Appl. Phys. Lett. {\bf{94}}, 092103 (2009).

\bibitem{Pierre.apl.2009} M. Pierre, R. Wacquez, B. Roche, X. Jehl, M. Sanquer, M. Vinet, E. Prati, M. Belli, and M. Fanciulli, Appl. Phys. Lett. {\bf{95}}, 242107 (2009).

\bibitem{Skinner.prl.2003} A. J. Skinner, M. E. Davenport, and B. E. Kane, Phys. Rev. Lett. {\bf{90}}, 087901 (2003).

\bibitem{Schofield.prl.2003} S. R. Schofield, N. J. Curson, M. Y. Simmons, F. J. Rue§, T. Hallam, L. Oberbeck, and R. G. Clark , Phys. Rev. Lett. {\bf{91}}, 136104 (2003).

%1D wire
\bibitem{Ruess.prb.2007.1} Frank J. Ruess, Bent Weber, Kuan Eng J. Goh, Oleh Klochan, Alex R. Hamilton, and Michelle Y. Simmons, Phys. Rev. B {\bf{76}}, 085403 (2007).

% 2D donor layer
\bibitem{Ruess.prb.2007} F. J. Ruess, W. Pok, K. E. J. Goh, A. R. Hamilton, and M. Y. Simmons, Phys. Rev. B {\bf{75}}, 121303(R) (2007).


\bibitem{Simmons.unpublished} Simmons group unpublished work (2009-2010).

\bibitem{Simmons.nnano.2010} Martin Fuechsle, S. Mahapatra, F. A. Zwanenburg, Mark Friesen, M. A. Eriksson, and Michelle Y. Simmons, doi:10.1038/nnano.2010.95.

\bibitem{Jamieson.apl.2005} D. N. Jamieson, C. Yang, T. Hopf, S. M. Hearne, C. I. Pakes, S. Prawer, M. Mitic, E. Gauja, S. E. Andresen, F. E. Hudson, A. S. Dzurak, and R. G. Clark, Appl. Phys. Lett. {\bf{86}}, 202101 (2005). 

\bibitem{Donkelaar.njp.2010} Jessica A Van Donkelaar, Andrew D Greentree, Andrew D C Alves, Lenneke M Jong, Lloyd C L Hollenberg, and David N Jamieson, New Journal of Physics {\bf{12}}, (2010).

\bibitem{Koiller.prl.2001} B. Koiller, X. Hu, and S. Das Sarma, Phys. Rev. Lett. {\bf{88}}, 027903 (2001). 

% Charge Qubit - Xuedong Hu
\bibitem{Hu.prb.2005} X. Hu, Belita Koiller, and S. Das Sarma, Phys. Rev. B {\bf{71}}, 235332 (2005). 

\bibitem{Cole.pra.2005} Jared H. Cole, Sonia G. Schirmer, Andrew D. Greentree, Cameron J. Wellard, Daniel K. L. Oi, and Lloyd C. L. Hollenberg, Phys. Rev. A {\bf{71}}, 062312 (2005).

\bibitem{Cole.pra.2006} Jared H. Cole, Andrew D. Greentree, Daniel K. L. Oi, Sonia G. Schirmer, Cameron J. Wellard, and Lloyd C. L. Hollenberg, Phys. Rev. A {\bf{73}}, 062333 (2006).

\bibitem{Testolin.pra.2007} M. J. Testolin, C. D. Hill, C. J. Wellard, and L. C. L. Hollenberg, Phys. Rev. A {\bf{76}}, 012302 (2007).

\bibitem{Rahman.prb.2009} R. Rahman, S. H. Park, J. H. Cole, A. D. Greentree, R. P. Muller, G. Klimeck, and L. C. L. Hollenberg, Phys. Rev. B {\bf{80}}, 035302 (2009). 

\bibitem{Slater.physrev.1954} J. C. Slater and G.F. Koster, Phys. Rev. Vol. {\bf{94}}, No. 6 (1954). 

\bibitem{Rahman.prl.2007} R. Rahman, C. J. Wellard, F. R. Bradbury, M. Prada, J. H. Cole, G. Klimeck, and L. C. L. Hollenberg, Phys. Rev. Lett. {\bf{99}}, 036403 (2007). 

\bibitem{Park.prl.2009} S. H. Park, R. Rahman, G. Klimeck, and L. C. L. Hollenberg, Phys. Rev. Lett. {\bf{103}}, 106802 (2009). 

\bibitem{Rogge.NaturePhysics.2008} G. P. Lansbergen, R. Rahman, C. J. Wellard, I. Woo, J. Caro, N. Collaert, S. Biesemans, G. Klimeck, L. C. L. Hollenberg, and S. Rogge, Nature Physics {\bf{4}}, 656 (2008).

\bibitem{Rahman.prb.2009.1} R. Rahman, G. P. Lansbergen, S. H. Park, J. Verduijn, G. Klimeck, S. Rogge, and L. C. L. Hollenberg, Phys. Rev. B {\bf{80}}, 165314 (2009).

\bibitem{Klimeck.cmes.2002} G. Klimeck, F. Oyafuso, T. B. Boykin, R. C. Bowen, and P. V. Allmen, Comput. Model. Eng. Sci. {\bf{3}}, 601, (2002).  

\bibitem{Klimeck.ted.2007} G. Klimeck, S. Ahmed, N. Kharche, M. Korkusinski, M. Usman, M. Prada, and T. Boykin, IEEE Trans. Electron Dev. {\bf{54}}, 2079 (2007).

%\bibitem{Tom.pra.2009} Tom\'a\^s Opatrn\'y and Kunal K. Das, Phys. Rev. A {\bf{79}}, 012113 (2009).

\bibitem{Kamleitner.pra.2008} I. Kamleitner, J. Cresser, and J. Twamley, Phys. Rev. A {\bf{77}}, 032331 (2008).

\bibitem{Greentree.spie.2005} A. D. Greentree, J. H. Cole, A. R. Hamilton, and L. C. L. Hollenberg, {\it{Micro- and Nanotechnology: Materials, Processes, Packaging and Systems II}} (SPIE-The International Society of Optical Engineers, University of NSW, Sydney, 2005), Vol. 5650, pp. 72-80.

\bibitem{Boykin.prb.2004} T. B. Boykin, G. Klimeck, and F. Oyafuso, Phys. Rev. B {\bf{69}}, 115201 (2004).

 \bibitem{Neerav.apl.2007} N. Kharche, Marta Prada, T. B. Boykin, G. Klimeck, Appl. Phys. Lett. {\bf{90}}, 092109 (2007).
 
\bibitem{Shaikh.encyclopedia.2008} S. Ahmed, N. Kharche, R. Rahman, M. Usman, S. Lee, H. Ryu, H. Bae, S. Clark, B. Haley, M. Naumov, F. Saied, M. Korkusinski, R. Kennel, M. McLennan, T. B. Boykin, and G. Klimeck, Encyclopedia of Complexity of Systems (2008).

\bibitem{ISE.TCAD} The electrostatic gate modeling was done using ISE TCAD in CQCT, Melbourne, Australia.

\bibitem{Lee.prb.2004} S. Lee, F. Oyafuso, P. von Allmen, and G. Klimeck, Phys. Rev. B {\bf{69}}, 045316 (2004).

\bibitem{nanohub.note} nanoHUB.org computational resource of a 256-node 3.3GHz Pentium Irvindale PC cluster was used in this work. The tight-binding calculations were done using the Nano-Electronic Modeling Tool (NEMO-3D). 

\bibitem{Ramdas.progphys.1981} A. K. Ramdas and S. Rodriguez, Rep. Prog. Phys., Vol. {\bf{44}} (1981).

\bibitem{Kohn.physrev.1955} W. Kohn and J. M. Luttinger, Phys. Rev. {\bf{98}}, 915 (1955).

 \end{thebibliography}
\end{document}